\documentclass[preprint,12pt]{elsarticle}
\usepackage{graphicx}
\usepackage{amssymb}
\usepackage{amsmath}
\usepackage{amsfonts}
\usepackage{slashed}
\usepackage[dvipsnames]{xcolor}
\def\lamb#1#2{$^{#1}_{\Lambda}${#2}}
\def\lam#1#2{$^{#1}_{~\Lambda}${#2}}

\journal{Physics Letters B}

\begin{document}

\begin{frontmatter}

\title{Has J-PARC E07 observed a $\Xi^-_{1s}$ nuclear state?}
\author{E.~Friedman} 
\author{A.~Gal} 
\address{Racah Institute of Physics, The Hebrew University, 9190401
Jerusalem, Israel} 

\begin{abstract} 
The Strangeness ${\cal S}=-2$ J-PARC E07 emulsion experiment published 
recently two new $\Xi^-$--$^{14}$N capture events, IRRAWADDY and IBUKI, 
identified by observing weak-decay sequences of pairs of single-$\Lambda$ 
hypernuclei and interpreted as $\Xi^{-}_{1s}$ and $\Xi^{-}_{1p}$ nuclear bound 
states, respectively, with binding energies $B_{\Xi^-}^{1s}$=6.27$\pm$0.27~MeV 
and $B_{\Xi^-}^{1p}$=1.27$\pm$0.21~MeV. $\Xi^-$ capture events in emulsion 
play a major role in determining the $\Xi$-nuclear and the $\Lambda\Lambda$ 
potential strengths. Here we question the assignment of a $\Xi^{-}_{1s}$ 
nuclear bound state to IRRAWADDY and offer an alternative assignment as a 
near-threshold $\Xi^0_{1p}$--$^{14}$C nuclear bound state, slightly admixed 
with a $\Xi^{-}_{1p}$--$^{14}$N nuclear bound state component corresponding 
to IBUKI. We also question using IBUKI's $B_{\Xi^-}^{1p}$ value as is to 
determine the $\Xi$-nuclear potential depth. Altogether, $\Xi^-$ capture 
events in $^{12}$C and $^{16}$O should prove less ambiguous than in $^{14}$N. 
\end{abstract} 


\end{frontmatter}

\section{Introduction} 
\label{sec:intro} 

Experimental input to deciphering the Strangeness ${\cal S}=-2$ sector 
of hypernuclear physics comes mostly from forming $\Xi^-$ hyperons by 
a $(K^-,K^+)$ reaction on nuclear targets and stopping them in light-nuclei 
emulsion~\cite{HN18}. Data collected in the last 30 years by KEK emulsion 
experiments E176~\cite{E176} and E373~\cite{E373,Nakazawa15} have advanced our 
understanding of $\Lambda\Lambda$ and $\Xi N$ nuclear dynamics~\cite{GHM16}. 
New J-PARC E07 $\Xi^-$ capture data, some of which were presented 
recently~\cite{Ekawa19,Hayakawa21,Yoshimoto21}, may advance it further. 

Because of the large momentum transfer in the $(K^-,K^+)$ nuclear reaction 
which converts protons to $\Xi^-$ hyperons, only very few of the produced 
high-energy $\Xi^-$ hyperons slow down sufficiently in the surrounding nuclear 
emulsion to undergo Auger process, form high-$n$ atomic state, and cascade 
down radiatively. In light-nuclei emulsion, strong-interaction $\Xi^- p
\to\Lambda\Lambda$ capture takes over radiative cascade beginning at a $3D$ 
atomic state bound by 126, 175, 231 keV in C, N, O, respectively, although the 
$3D$ strong-interaction shift is still small, less than 1~keV~\cite{BFG99}. 
Capture events are recorded by observing $\Lambda$ hyperons or 
$\Lambda$-hypernuclear decay products. Interestingly, several observed decay 
sequences of twin (pairs of) single-$\Lambda$ hypernuclei were found to follow 
capture from a lower $\Xi^-$ orbit, namely Coulomb-assisted $1p_{\Xi^-}$ 
nuclear states in $^{12}$C and $^{14}$N~\cite{E176,Nakazawa15,Hayakawa21}. 

\begin{table}[htb]
\begin{center}
\caption{Reported two-body $\Xi^-$ capture events in $^{14}$N, 
$\Xi^-$+$^{14}$N $\to$~\lam{A'}{Z'}+\lam{A''}{Z''}, to twin $\Lambda$ 
hypernuclei, some in ground states, some in specific excited states marked 
by asterisk. Of these, only IBUKI is uniquely assigned. Fitted $\Xi^-$ 
binding energies $B_{\Xi^-}^{1p}$ are listed.}
\begin{tabular}{cccc}
\hline 
Experiment & Event & \lamb{A'}{Z'}+\lamb{A''}{Z''} & $B_{\Xi^-}^{1p}$
(MeV)  \\
\hline 
KEK E176 \cite{E176} & 14-03-35 & \lamb{3}{H}+\lam{12}{B} &
1.18$\pm$0.22  \\
KEK E373 \cite{Nakazawa15} & KISO & \lamb{5}{He}+\lam{10}{Be}$^{\ast}$ &
1.03$\pm$0.18  \\
J-PARC E07 \cite{Hayakawa21} & IBUKI & \lamb{5}{He}+\lam{10}{Be} &
1.27$\pm$0.21  \\
\hline
\end{tabular}
\label{tab:14N1p}
\end{center}
\end{table}

The J-PARC light-nuclei emulsion experiment E07 reported two $\Xi^-$ capture 
events in $^{14}$N, uniquely identified by their subsequent single-$\Lambda$ 
hypernuclear decay sequences, and with sufficiently small binding-energy 
uncertainty (say $\lesssim 0.3$~MeV) to become potentially informative. 
One event, named IBUKI, was assigned as a $\Xi^-_{1p}$--$^{14}$N bound state, 
with binding energy $B_{\Xi^-}$ listed in Table~\ref{tab:14N1p} together with 
similar $B_{\Xi^-}$ values for $\Xi^-$--$^{14}$N capture events found in KEK 
experiments E176 and E373. All three listed $B_{\Xi^-}$ values are slightly 
above 1~MeV, significantly higher than the purely-Coulomb $2P$ atomic 
binding energy value of about 0.4 MeV in $^{14}$N, thereby corresponding to 
a Coulomb-assisted $1p_{\Xi^-}$ nuclear state in $^{14}$N that evolves from a 
$2P$ atomic state. Unfortunately, it is just one of five $\Xi^-_{1p}$--$^{14}
$N$_{\rm g.s.}$ configuration states spread over $\sim$1.3~MeV (see Appendix 
to the present work) which makes IBUKI unfit to constrain reliably the 
$\Xi$-nuclear potential strength~\cite{FG21}. 

\begin{figure}[!t] 
\begin{center} 
\includegraphics[width=0.8\textwidth]{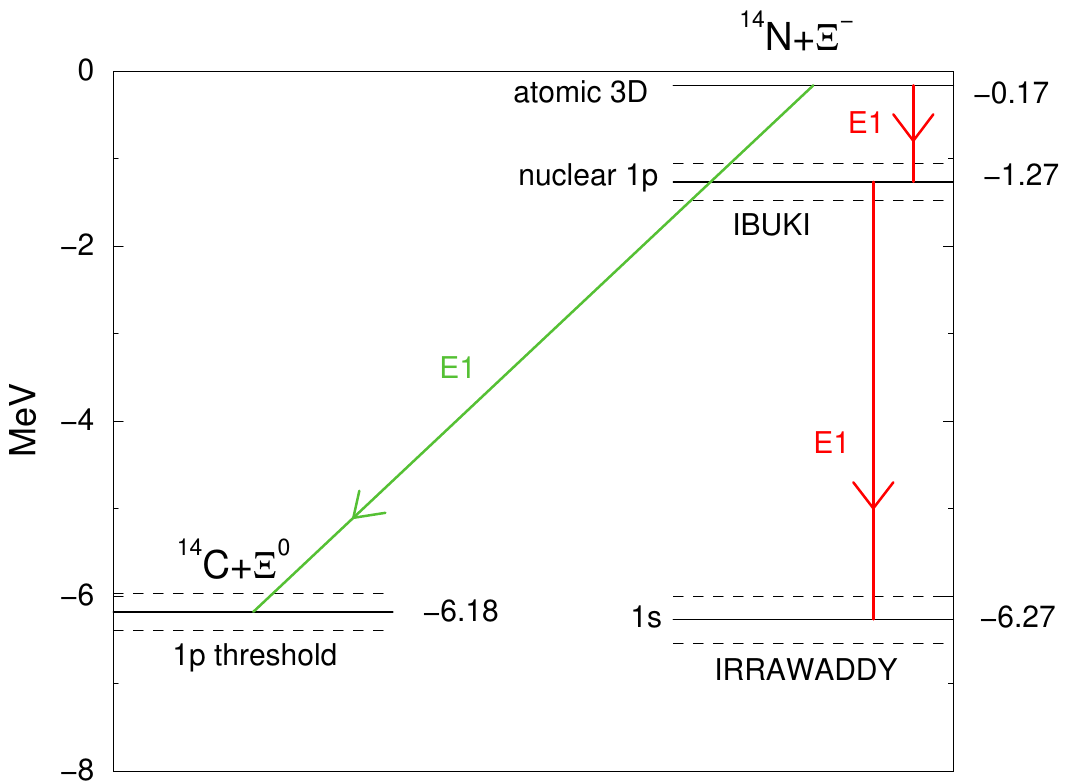} 
\caption{Level diagram of $\Xi^-$+$^{14}$N. Shown on the right are a $3D$ 
atomic state and $1p,1s$ nuclear states assigned, respectively, to E07 
$\Xi^-$ capture events IBUKI and IRRAWADDY~\cite{Hayakawa21,Yoshimoto21}. 
The $\Xi^0$+$^{14}$C threshold at $-$6.18~MeV is marked on the left. 
Electromagnetic $E1$ transitions deexciting the $\Xi^-_{3D}$ atomic state to 
lower $\Xi^-$--$^{14}$N and $\Xi^0_{1p}$--$^{14}$C nuclear states from which 
$\Xi$ hyperon capture may occur are marked by red and green arrowed lines, 
respectively. A near-threshold $\Xi^0_{1p}$--$^{14}$C state on the left 
provides an alternative interpretation of IRRAWADDY.} 
\label{fig:Xiplot} 
\end{center} 
\end{figure} 

The other event, with binding energy $B_{\Xi^-}$=6.27$\pm$0.27~MeV, was named 
IRRAWADDY and assigned as a $\Xi^-_{1s}$ nuclear state~\cite{Yoshimoto21}. 
Both IBUKI and IRRAWADDY are shown on the r.h.s. of Fig.~\ref{fig:Xiplot}, 
fed by radiative E1 transitions from the upper 3$D$ $\Xi^-$-atom state. 
Another $\Xi^-_{1s}$ candidate event KINKA, with two possible $B_{\Xi^-}$ 
values of 8.00$\pm$0.77 and 4.96$\pm$0.77~MeV, was deduced by revisiting 
KEK-E373 capture events~\cite{Yoshimoto21}. A KINKA+IRRAWADDY weighted 
average of $B_{\Xi^-}$=6.13$\pm$0.25 or 6.46$\pm$0.25~MeV, differs little 
from IRRAWADDY's own value of $B_{\Xi^-}$=6.27$\pm$0.27~MeV. 

We note that $2P\to 1S$ radiative decay rates are of order 1\% of $3D\to 2P$ 
radiative decay rates~\cite{Zhu91,Koike17} suggesting that $\Xi^-$ capture 
from a nuclear $\Xi^-_{1s}$--$^{14}$N state is suppressed to this order 
relative to capture from a nuclear $\Xi^-_{1p}$--$^{14}$N state. 
Assigning a $\Xi^-_{1s}$--$^{14}$N bound state to IRRAWADDY is therefore 
questionable. Here we explore an alternative assignment, as follows. 

Little attention if any was given to mixing between $\Xi^-$ states in $^{14}$N 
and $\Xi^0$ states in $^{14}$C, induced by the  $\Xi^- p\leftrightarrow
\Xi^0 n$ strong-interaction charge exchange. Nuclear ground-state mass 
differences for $Z\to (Z-1)$ isobars in the main light-nuclei emulsion 
elements $^{12}$C, $^{14}$N and $^{16}$O are listed in Table~\ref{tab:cex}. 
Note that the nuclear mass difference $M(^{14}{\rm C})-M(^{14}{\rm N})$ 
is smaller than the $\Xi$ hyperon mass difference $m_{\Xi^-}-m_{\Xi^0}=6.85 
\pm 0.21$~MeV \cite{pdg22}. This means that $\Xi^0$--$^{14}$C bound states, 
if and when exist, lie always below the $\Xi^-$--$^{14}$N threshold. 
To be more precise, the $\Xi^0$--$^{14}$C threshold for a virtual 
$^{14}$N($\Xi^-,\Xi^0$)$^{14}$C charge-exchange reaction is at 
$\Xi^-$--$^{14}$N energy of 
\begin{equation} 
-(6.85\pm 0.21)+0.67 = -(6.18\pm 0.21)~{\rm MeV}. 
\label{eq:cex} 
\end{equation}  
Remarkably, the $\Xi^0$--$^{14}$C threshold energy in $\Xi^-$--$^{14}$N, 
Eq.~(\ref{eq:cex}), is within the reported IRRAWADDY bound-state 
energy interval $-$(6.27$\pm$0.27) MeV, as shown in Fig.~\ref{fig:Xiplot}.  

\begin{table}[!h]
\begin{center}
\caption{Ground-state nuclear mass differences (in MeV) for $Z\to Z-1$ isobars 
of the main light-nuclei emulsion elements $^{12}$C, $^{14}$N and $^{16}$O 
\cite{CP22}.} 
\begin{tabular}{ccc}
\hline
$M(^{12}$B)--$M(^{12}$C) & $M(^{14}$C)--$M(^{14}$N) & $M(^{16}$N)--$M(^{16}$O) 
 \\
\hline 
13.88 & 0.67 & 10.93  \\ 
\hline 
\end{tabular} 
\label{tab:cex} 
\end{center} 
\end{table} 

The aim of the present note is to show that this remarkable coincidence may 
not be accidental, and that IRRAWADDY is likely to stand for a near-threshold 
$\Xi^0_{1p}$--$^{14}$C bound state that has nothing to do with a $\Xi^-_{1s}
$--$^{14}$N bound state suggested by E07. To do so, we first discuss in 
Sect.~2 conditions for a near-threshold $\Xi^0_{1p}$--$^{14}$C bound state 
using a $\Xi$-nuclear potential constrained by our recent work~\cite{FG21} on 
$\Xi^-_{1p}$--$^{12}$C capture events. We then evaluate in Sect.~3 the mixing 
induced by the $\Xi N$ strong interaction between a $\Xi^-_{1p}$--$^{14}$N 
bound state identified with J-PARC E07 IBUKI and a $\Xi^0_{1p}$--$^{14}$C 
bound state lying about 5 MeV below IBUKI, within the J-PARC E07 experimental 
uncertainty of IRRAWADDY. Provided such mixing is not prohibitively small, 
$E1$ radiative deexcitation of the $\Xi^-_{3D}$--$^{14}$N atomic state 
occurs to {\it both} $\Xi^-_{1p}$--$^{14}$N and $\Xi^0_{1p}$--$^{14}$C 
nuclear states, as verified in Sect.~3. Discussion and summary follow in 
Sect.~4 and a brief Appendix describes the $\Xi^-_{1p}$--$^{14}$N$_{g.s.}$ 
configuration.

\section{$\Xi^0_{1p}$--$^{14}$C binding} 
\label{sec:bind} 

Following the optical model methodology introduced in our recent work on 
low-energy $\Xi^-$--nuclear interactions~\cite{FG21}, the strong-interaction 
optical potential used here is of the form 
\begin{equation} 
V_{\rm opt}^{\Xi}(r)=-\frac{2\pi}{\mu}(1+\frac{A-1}{A}\frac{\mu}{m_N})
[b_0(\rho)\,\rho(r)+b_1\,\rho_{\rm exc}(r)]. 
\label{eq:EEs} 
\end{equation} 
Here $\mu$ is the $\Xi^-$-nuclear reduced mass, and $b_0$ and $b_1$ are  
effective $\Xi N$ c.m. scattering amplitudes related to the isoscalar $V_0$ 
and isovector $V_{\tau}$ components, respectively, of the two-body $s$-wave 
$\Xi N$ interaction $V_{\Xi N}$,
\begin{equation} 
V_{\Xi N}=V_0+V_{\sigma}{\vec\sigma}_{\Xi}\cdot{\vec\sigma}_N+V_{\tau}
{\vec\tau}_{\Xi}\cdot{\vec\tau}_N+V_{\sigma\tau}{\vec\sigma}_{\Xi}\cdot
{\vec\sigma}_N\,{\vec\tau}_{\Xi}\cdot{\vec\tau}_N \, ,
\label{eq:V2body} 
\end{equation}
with $V$s functions of $r_{\Xi N}$. The density $\rho=\rho_n+\rho_p$ in 
Eq.~(\ref{eq:EEs}) is a nuclear density distribution normalized to $A$, 
and $\rho_{\rm exc}=\rho_n-\rho_p$ is a neutron-excess density with 
$\rho_n=(N/Z)\rho_p$, implying that $\rho_{\rm exc}=0$ for the $N=Z$ emulsion 
nuclei $^{12}$C, $^{14}$N and $^{16}$O. The r.m.s. radius of $\rho_p$ was 
set equal to that of known nuclear charge densities, as discussed in our 
earlier work~\cite{FG21}. Finally, the density-dependent $b_0(\rho)$ is 
given by 
\begin{equation} 
b_0(\rho)=\frac{{\rm Re}\,b_0}{1+\frac{3k_F}{2\pi}{\rm Re}\,b_0^{\rm lab}}+
{\rm Im}\,b_0, \,\,\,\,\,\,\,\,\,\,k_F=(3{\pi}^2\rho/2)^{\frac{1}{3}}, 
\label{eq:WRW} 
\end{equation} 
with Fermi momentum $k_F$ and $b_0^{\rm lab}=(1+\frac{m_{\Xi^-}}{m_N})b_0$. 
Eq.~(\ref{eq:WRW}) accounts for Pauli correlations in $\Xi N$ in-medium 
multiple scatterings~\cite{WRW97}. The value of Im$\,b_0$ is fixed at 0.01~fm 
to simulate $\Xi N\to\Lambda\Lambda$ conversion. 

Using $b_0=(0.527+{\rm i}0.010)$~fm from fitting $B^{1p}_{\Xi^-}=0.82$~MeV in 
$^{12}$C~\cite{FG21}, together with $b_1=0$, failed to produce a $\Xi^0_{1p}
$--$^{14}$C bound state. This agrees with $\Xi^-_{1p}$--$^{12}$C being 
a Coulomb-assisted bound state, unbound without the $\Xi^-$-nuclear 
Coulomb potential $V_C$. However, a likely source of additional attraction 
for $\Xi^0_{1p}$--$^{14}$C, with $N\neq Z$, is provided by the $\Xi N$ 
strong-interaction $V_{\tau}$ isovector component in terms of $b_1\neq 0$. 
Indeed, varying $b_1$ between 0.270 to 0.420~fm produces $\Xi^0_{1p}$ 
binding energies in $^{14}$C ranging between $-$200 to 178 keV, comfortably 
within IRRAWADDY's binding energy uncertainty, and widths between 274 to 
409~keV.{\footnote{Normalizable $1p$ bound states above the $\Xi^0$--$^{14}$C 
threshold are generated owing to the imaginary part of $V_{\rm opt}^{\Xi}$ in 
Eq.~(\ref{eq:EEs}), see Ref.~\cite{GTA81} for discussion of $\Sigma$ nuclear 
states above threshold.}} These $b_1$ values, relative to $b_0$, are of the 
same sign and order of magnitude expected from the HAL-QCD lattice derivation 
of $V_{\Xi N}$~\cite{Sasaki20} and from the chiral EFT calculation at NLO 
of $\Xi$ in nuclear matter~\cite{HM19}. The corresponding $\Xi^0_{1p}$ 
radial wavefunctions reach a maximum value between 2.66 to 2.54~fm in 
the 378~keV binding energy interval specified above. Furthermore, the 
$\Xi^-_{1p}$--$^{14}$N Coulomb-assisted bound state wavefunction generated 
for the same value of $b_0$, but $b_1=0$ as appropriate to the $T=0$ $^{14}$N, 
reaches its maximum at 2.62~fm, overlapping to better than 0.98 with any of 
the $\Xi^0_{1p}$--$^{14}$C wavefunctions in the range considered above.

\section{$\Xi^0_{1p}$--$^{14}$C $\leftrightarrow$ $\Xi^-_{1p}$--$^{14}$N 
mixing} 
\label{sec:mix} 

To evaluate the $\Xi^-_{1p}$--$^{14}$N to $\Xi^0_{1p}$--$^{14}$C 
mixing amplitude we use realistic shell-model wavefunctions for 
$^{14}$N$_{\rm g.s.}$ and $^{14}$C$_{\rm g.s.}$~\cite{Mill07}: 
\begin{equation} 
|{^{14}{\rm C}}(J^{\pi}=0^+;T=1)\rangle = 0.7729\,{^{1}S}+0.6346\,{^{3}P}\, , 
\label{eq:14C} 
\end{equation} 
\begin{equation} 
|{^{14}{\rm N}}(J^{\pi}=1^+;T=0)\rangle = 0.1139\,{^{3}S} + 0.2405\,{^{1}P} 
+ 0.9639\,{^{3}D} \, , 
\label{eq:14N} 
\end{equation} 
where the $^{2S+1}L$ spin-orbital basis wavefunctions refer to 
the two-hole $(1p_N)^{-2}$ configuration in the nuclear $p$-shell. 
A check on the $S$ and $P$ amplitudes is provided by the near vanishing 
of the weak-interaction Gamow-Teller (GT) $\Delta S$=$\Delta T$=1 matrix 
element $\langle{^{14}{\rm N}}(1^+;0)|{\sum \vec{\sigma}_i \tau^{+}_i}|
{^{14}{\rm C}}(0^+;1)\rangle$ responsible for the extremely long lifetime 
of $^{14}$C$_{\rm g.s.}(0^+;1)$ in its $\beta$-decay to $^{14}$N$_{\rm g.s.}
(1^+;0)$. The near vanishing of this GT matrix element occurs owing to 
a strong cancelation between the $\Delta L$=0 $^3S_1\leftrightarrow{^1{S_0}}$ 
and $^1P_1\leftrightarrow {^3{P_0}}$ GT transitions, persisting also 
for the same $\Delta L$=0 transitions induced by the $V_{\sigma\tau}
{\vec\sigma}_{\Xi}\cdot{\vec\sigma}_N\,{\vec\tau}_{\Xi}\cdot{\vec\tau}_N$ 
term of the strong-interaction $V_{\Xi N}$, Eq.~(\ref{eq:V2body}), in mixing 
$\Xi^-_{1s}$--$^{14}$N$(1^+;0)$ with $\Xi^0_{1s}$--$^{14}$C$(0^+;1)$. However, 
the $\Xi^-_{1p}$--$^{14}$N$(1^+;0)$ to $\Xi^0_{1p}$--$^{14}$C$(0^+;1)$ mixing 
is governed by a normal-strength $\Delta L$=2 matrix element that couples the 
dominant $^3D_1$ amplitude in $^{14}$N$_{\rm g.s.}(1^+;0)$ to the $^1S_0$ and 
$^3P_0$ amplitudes in $^{14}$C$_{\rm g.s.}(0^+;1)$. 

\begin{table}[htb]
\begin{center}
\caption{Coefficients of $F^{(2)}_{\Xi N}(V_{\sigma\tau})$ in 
mixing matrix elements between two $J^{\pi}={\frac{1}{2}}^-$ states: 
$\Xi^0_{1p}$--$^{14}$C$(0^+;1)$ and the $\Xi^-_{1p}$--$^{14}$N$(1^+;0)$ 
IBUKI state. Entries are labeled by $^{2S+1}L$ components of $A=14$ 
wavefunctions given by Eqs.~(\ref{eq:14C}) and (\ref{eq:14N}).}
\begin{tabular}{cccc}
\hline
$^{2S+1}L$ &~~~$^{3}S$~~~&~~~$^{1}P$~~~&~~~$^{3}D$~~~  \\ 
\hline 
$^{1}S$ & -- & -- & $\frac{4{\sqrt 2}}{5{\sqrt 5}}$  \\ 
$^{3}P$ & -- & $\frac{2{\sqrt 2}}{5{\sqrt 3}}$ & $\frac{6}{5{\sqrt 5}}$  \\ 
\hline
\end{tabular}
\label{tab:mix}
\end{center}
\end{table}

The IBUKI state, as detailed in the Appendix, is a $J^{\pi}={\frac{1}{2}}^-$ 
state built by coupling a $s_{\Xi^-}=\frac{1}{2}$ Pauli-spin to 
${\Lambda}^{\pi}=0^-$, where $\vec{\Lambda}={\vec J}_N+{\vec \ell}_{\Xi}$. 
The full $\Xi^-_{1p}$--$^{14}$N$(1^+;0)$ configuration includes also 
${\Lambda}^{\pi}=1^-,2^-$ states. As for the $\Xi^0_{1p}$--$^{14}$C$(0^+;1)$ 
configuration, it consists of a $J^{\pi}=({\frac{1}{2}}^-,{\frac{3}{2}}^-)$ 
spin-orbit doublet obtained by coupling $s_{\Xi^0}=\frac{1}{2}$ to the only 
possible ${\Lambda}^{\pi}=1^-$ state. Mixing matrix elements of $V_{\sigma\tau}
{\vec\sigma}_{\Xi}\cdot{\vec\sigma}_N\,{\vec\tau}_{\Xi}\cdot{\vec\tau}_N$ 
between these $J^{\pi}={\frac{1}{2}}^-$ states are listed in 
Table~\ref{tab:mix} in terms of $LS$ basis components of the $^{14}$C$(0^+;1)$ 
and $^{14}$N$(1^+;0)$ ground-state wavefunctions, Eqs.~(\ref{eq:14C}) and 
(\ref{eq:14N}). These matrix elements need to be multiplied by the 
corresponding Slater integral $F^{(2)}_{\Xi N}(V_{\sigma\tau})$. 

Using Eqs.~(\ref{eq:14C}) and (\ref{eq:14N}) for the nuclear $LS$ basis 
amplitudes and the $LS$ matrix elements listed in Table~\ref{tab:mix}, 
one gets the mixing matrix element between the $J^{\pi}={\frac{1}{2}}^-$ 
relevant $\Xi^0_{1p}$--$^{14}$C$_{\rm g.s.}$ and $\Xi^-_{1p}$--$^{14}
$N$_{\rm g.s.}$ states in terms of $F^{(2)}_{\Xi N}(V_{\sigma\tau})$: 
\begin{equation} 
\langle \Xi^0_{1p}-^{14}{\rm C}_{\rm g.s.}|V_{\sigma\tau}{\vec\sigma}_{\Xi}
\cdot{\vec\sigma}_N\,{\vec\tau}_{\Xi}\cdot{\vec\tau}_N|\Xi^-_{1p}-^{14}
{\rm N}_{\rm g.s.}\rangle =(0.705+0.050)F^{(2)}_{\Xi N}(V_{\sigma\tau}). 
\label{eq:mix} 
\end{equation} 
Here the two matrix elements involving the dominant $^{3}D$ component in 
$^{14}$N$(1^+;0)$ contribute coherently to give most of the mixing, 0.705$
F^{(2)}$, and the contribution of the $P$-states mixing, 0.050$F^{(2)}$, is 
small on this scale. For a conservative estimate we take $F^{(2)}_{\Xi N}
(V_{\sigma\tau})\approx -0.6$~MeV, which is 20\% of the value of $F^{(2)}_{
\Xi N}(V_0)$ used in Ref~\cite{FG21} to reproduce IBUKI, thereby getting 
$\approx\,-$0.453~MeV for the overall mixing matrix element. The resulting 
amplitude of IBUKI's admixture into $\Xi^0_{1p}$--$^{14}$C$_{\rm g.s.}$, now 
identified with IRRAWADDY, is obtained upon dividing it by $\Delta E_{1p} 
\approx 5$~MeV for the energy difference between the two admixed $1p_{\Xi}$ 
nuclear states: $-0.453/5 = -0.0906$, and 0.00821 upon squaring it. 
The radiative $E1$ transition rate from $3D_{\Xi^-}$ to $1p_{\Xi}$ 
states through their $\Xi^-_{1p}$--$^{14}$N$_{\rm g.s.}$ components goes 
like $(\Delta E)^3$, with a large factor of $(6.1/1.1)^3\approx 171$ in 
favor of populating the lower, $\Xi^0_{1p}$--$^{14}$C$_{\rm g.s.}$ state. 
Altogether one gets a ratio of 0.00821$\times$171=1.40 between the two $E1$ 
rates, making them roughly equal to each other. A schematic plot of these 
two $E1$ transitions is shown in Fig.~\ref{fig:Xiplot}.

\section{Discussion and summary} 
\label{sec:concl} 

Before summarizing it is appropriate to discuss briefly the role played by 
the choice of $\Xi^-$ capture data input in predicting $\Xi$-nuclear bound 
states and potential depths. In our recent work \cite{FG21} two KEK-E176 
$^{12}$C events \cite{E176} with $B_{\Xi^-}^{1p}(^{12}$C)=0.82$\pm$0.14~MeV 
served as input for setting up the $\Xi$-nuclear optical potential depth 
at nuclear-matter density, $-V_{\Xi}(\rho_0)$=21.9$\pm$0.7~MeV, and for 
explaining IBUKI's $B_{\Xi^-}^{1p}(^{14}$N)=1.27$\pm$0.21~MeV (see Appendix). 
Predicting $B_{\Xi^-}^{1s}$ values from a $B_{\Xi^-}^{1p}$ input value 
requires introduction of a more systematic density dependence, as done 
in $\Lambda$ hypernuclear studies~\cite{FG22}. Choosing alternatively 
the J-PARC E07 $^{14}$N IRRAWADDY event~\cite{Yoshimoto21} as input, 
with $B_{\Xi^-}^{1s}(^{14}$N)=6.27$\pm$0.27~MeV, gives rise to predictions 
listed in Table~\ref{tab:targets}. Among the $\Xi^-_{1p}$ bound states listed 
in the table, the $\Xi^-_{1p}$--$^{12}$C binding energy comes out exceedingly 
small compared to that determined from the KEK E176~\cite{E176} capture 
events.{\footnote{Regarding $\Xi^-_{1p}$--$^{12}$C, the density-folded 
$G$-matrix Ehime potential calculation of Ref.~\cite{Ehime01}, using 
an outdated value $B_{\Xi^-}^{1p}(^{12}$C)=0.58~MeV for input, nearly 
reproduced IRRAWADDY and IBUKI long time before their observation. 
Recalling, however, that 0.58~MeV is almost half the way from the KEK 
E176 reported value 0.82~MeV~\cite{E176} down to the pure-Coulomb value 
$B_{\Xi^-}^{2P}(^{12}$C)=0.28~MeV, their results require revision. 
Furthermore, since $J({^{14}{\rm N}})\neq 0$ (see Appendix) their calculated 
$B_{\Xi^-}^{1p}(^{14}$N) value cannot be compared directly with IBUKI's 
binding energy of 1.27$\pm$0.27~MeV in $^{14}$N~\cite{Hayakawa21}.}} We note 
that a 13.8~MeV potential depth value listed in the table agrees closely with 
the depth derived in a recent Skyrme-Hartree-Fock calculation~\cite{GZS21} 
that focuses on the $^{14}$N KINKA and IBUKI capture events. 

\begin{table}[!h] 
\begin{center} 
\caption{$\Xi^-$--$^{12}$C and $\Xi^-$--$^{14}$N binding energies (MeV) 
in $1s$ and $1p$ states, plus the $\Xi$ nuclear potential depth (MeV) 
at nuclear-matter density $\rho_0=0.17\,$fm$^{-3}$, calculated using a  
density-dependent optical potential Eqs.~(\ref{eq:EEs},\ref{eq:WRW}) with 
Re$\,b_0$ fitted to the $\Xi^-$--$^{14}$N $1s$ binding energy assigned 
by J-PARC E07~\cite{Yoshimoto21} to IRRAWADDY as underlined here.} 
\begin{tabular}{cccccc} 
\hline 
Input & $\Xi^-_{1s}$--$^{12}$C & $\Xi^-_{1p}$--$^{12}$C 
 & $\Xi^-_{1s}$--$^{14}$N & $\Xi^-_{1p}$--$^{14}$N & $-V_{\Xi}(\rho_0)$  \\ 
\hline 
IRRAWADDY & 4.94 & 0.32 & $\underline{6.27}$ & 0.50 & 13.8  \\ 
\hline 
\end{tabular} 
\label{tab:targets} 
\end{center} 
\end{table} 

The solution offered here to the difficulty of interpreting IRRAWADDY as 
a $\Xi^-_{1s}$ bound state in $^{14}$N is by pointing out that it could 
correspond to a $\Xi^0_{1p}$--$^{14}$C bound state, something that cannot 
happen kinematically in the other light-emulsion nuclei $^{12}$C and $^{16}$O. 
We showed in the present work that radiative $E1$ deexcitation of the 
$\Xi^-$--$^{14}$N atom, proceeding through the $\Xi^-_{3D}$ atomic state, 
leads at comparable rates to two $\Xi_{1p}$-nuclear states coupled by 
$\Xi^-p \leftrightarrow\Xi^0n$ strong-interaction charge exchange. One state, 
of dominantly $\Xi^-_{1p}$--$^{14}$N structure is the one assigned commonly 
to IBUKI, and the other one, of dominantly $\Xi^0_{1p}$--$^{14}$C structure 
is assigned here to IRRAWADDY. The latter assignment contrasts with viewing 
IRRAWADDY as a $\Xi^-_{1s}$--$^{14}$N state, an assignment motivated only 
by its binding energy of a few MeV. Given that in this nuclear mass range 
capture rates from $1s_{\Xi^-}$ states are estimated to be two orders of 
magnitude below capture rates from $1p_{\Xi^-}$ states~\cite{Zhu91,Koike17}, 
this $\Xi^0_{1p}$--$^{14}$C new assignment addresses satisfactorily the 
capture rate hierarchy. Establishing $1s_{\Xi^-}$ states in light emulsion 
nuclei requires focusing on capture events in $^{12}$C and $^{16}$O, where 
the $\Xi^-p\leftrightarrow\Xi^0n$ coupling is ineffective. 



\newpage 

\section*{Appendix A.~~$\Xi^-_{1p}$ spectrum in $^{14}$N}
\label{sec:appendix}
\renewcommand{\theequation}{A.\arabic{equation}}
\setcounter{equation}{0}
\renewcommand{\thefigure}{A.\arabic{figure}}
\setcounter{figure}{0}

\begin{figure}[h] 
\begin{center} 
\includegraphics[width=0.7\textwidth]{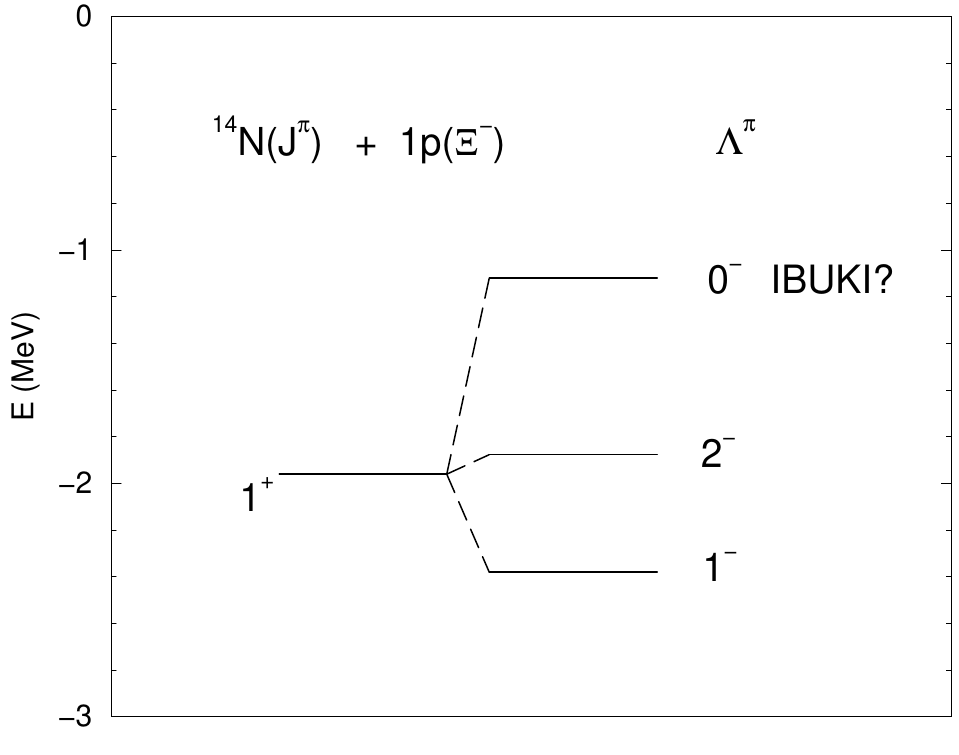} 
\caption{Spectrum of $^{14}$N$_{\rm g.s.} + 1p_{\Xi^-}$ states labeled 
by ${\Lambda}^{\pi}$, see text. Figure revising Fig.~2 in Ref.~\cite{FG21}.} 
\label{fig:14N} 
\end{center} 
\end{figure} 

The $\Xi^-_{1p}$--$^{14}$N$(1^+;0)$ configuration consists of two $J^{\pi}=
{\frac{1}{2}}^-$ states, two $J^{\pi}={\frac{3}{2}}^-$ states and one 
$J^{\pi}={\frac{5}{2}}^-$ state obtained by coupling $s_{\Xi^-}=\frac{1}{2}$ 
Pauli-spin to $\vec{\Lambda}={\vec J}_N+{\vec \ell}_{\Xi}$. Here $J_N^{\pi}=
1^+$, ${\ell}_{\Xi}^{\pi}=1^-$, so that $\Lambda^{\pi}=0^-,1^-,2^-$ as shown 
in Fig.~\ref{fig:14N}. The energy splittings marked in the figure follow from 
a shell-model quadrupole-quadrupole spin-independent residual interaction
${\cal V}_{\Xi N}^Q$,
\begin{equation} 
{\cal V}_{\Xi N}^Q=F^{(2)}_{\Xi N} Q_N\cdot Q_{\Xi}, \,\,\,\,\,\, 
Q_B=\sqrt{\frac{4\pi}{5}} Y_2({\hat{r}}_B), 
\label{eq:QdotQ}
\end{equation}
where $F^{(2)}$ is the corresponding Slater integral. For more details see 
Ref.~\cite{FG21} where it was also argued that the dependence on $s_{\Xi^-}$ 
is secondary to the dependence on $\Lambda$. The $(2{\Lambda}+1)$-averaged 
energy $-1.96\pm 0.26$~MeV was calculated using a central optical potential 
$V_{\rm opt}^{\Xi}$ specified in Sect.~\ref{sec:bind}, with $b_0$ fitted to 
KEK-E176 $\Xi^-$--$^{12}$C capture events~\cite{FG21}. 
The energy $E=-1.12$~MeV of the ${\Lambda}^{\pi}=0^-$ state is close to the 
energy $E\approx -1.16$~MeV deduced from the $\Xi^-$--$^{14}$N capture events 
listed in Table~\ref{tab:14N1p}. Thus, the IBUKI state corresponds to a unique 
$J^{\pi}={\frac{1}{2}}^-$ state built on ${\Lambda}^{\pi}=0^-$, the highest 
in the $\Lambda$ spectrum shown in the figure. To the best of our knowledge, 
this has not been realized in any past calculation of the $\Xi^-_{1p}$--$^{14}
$N$(1^+;0)$ spectrum. 

\section*{Acknowledgments} 

We thank Dr. K. Nakazawa for useful correspondence with one of us (A.G.) 
following his talk at HYP2022. This work was supported by the European 
Union's Horizon 2020 research and innovation programme under grant 
agreement No. 824093.

\end{document}